\documentclass[11pt,twoside]{article}

\usepackage{asp2004}
\usepackage{epsf}
\usepackage{psfig}
\usepackage{lscape}
\usepackage{amssymb}
\usepackage{graphicx}

\newcommand{\msun}{\ensuremath{\mathit{M}_{\odot}}}                  

\newcommand{\rg}{\mbox{$R_{\rm G}$}}

\newcommand{\rh}{\mbox{$r_{\rm h}$}}

\newcommand{\Mc}{\mbox{$M$}}

\newcommand{\tdis}{\mbox{$t_{\rm dis}$}}
\newcommand{\trel}{\mbox{$t_{\rm rh}$}}
\newcommand{\tcr}{\mbox{$t_{\rm cr}$}}

\newcommand{\rhoc}{\mbox{$\rho_{\rm c}$}}

\newcommand{\dr}{\mbox{${\mbox d}$}}

\newcommand{\dndm}{\mbox{$\dr N/\dr M$}}
\newcommand{\dndt}{\mbox{$\dr N/\dr t$}}

\begin{document}
\title{Conference summary: Mass loss from stellar clusters}
\author{M. Gieles}   
\affil{Astronomical Institute, Utrecht University, Princetonplein 5, 3584-CC Utrecht, The 
Netherlands}    
\affil{European Southern Observatory, Casilla 19001, Santiago 19, Chile}

\section{Introduction: topics on this conference}
This conference dealt with the mass loss from stars and from stellar
clusters. In this summary of the cluster section of the conference, I highlight some of the results on
the formation and the fundamental properties of star clusters (\S~\ref{sum:sec2}), the early stages of their evolution (\S~\ref{sum:sec3}) and go into more detail on the subsequent
mass evolution of clusters 
(\S~\ref{sum:sec4}). A discussion on  how this may, or may not, depend on mass is given in \S~\ref{sum:sec5} Obviously, there will be a bias towards the topics
where Henny Lamers has contributed. Some of the
contributions to these proceedings have already reviewed extensively the topics of
 clusters mass loss and disruption, so I will try to fit these in a general framework as much as possible.

\section{Cluster formation and fundamental properties}Ê
\label{sum:sec2}
Star clusters are considered to be the lowest step in the hierarchical star formation scenario (see Elmegreen in these proceedings). This hierarchy is thought to result from a combination of  (gravitational) fragmentation and turbulent compression in the interstellar medium (ISM), resulting in scale-free power spectra. These have been observed for the Milky Way ISM  \citep{2001ApJ...561..264D} and seem also to exist for young star fields in M33 \citep{2003ApJ...593..333E} and NGC~628 \citep{2006ApJ...644..879E}.
Fragmentation in a turbulent medium yields power-law distributions  at all scales for the mass and
radius distributions. Power-law mass functions are
found for the stellar mass function (e.g. \citealt{1955ApJ...121..161S}) and for the
cluster initial mass function (CIMF) \citep{1994A&AS..104..379B, 1999ApJ...527L..81Z, 2003MNRAS.343.1285D, 2003A&A...397..473B}, with some evidence for an upper-mass truncation (\citealt{2006A&A...450..129G, 2006A&A...446L...9G, 2006astro.ph.11586D} and Haas et al. in these proceedings). A truncation is also found for the mass function of giant molecular clouds \citep{1997ApJ...476..166W, 2005PASP..117.1403R}, but a solid theoretical understanding of the top end of the CIMF is still lacking. When extrapolating the cluster mass function to $N=1$, \citet{2005A&A...437..247D} show that 4\% of the O stars should form in isolation, which is in agreement with their observations. On a larger scale, complexes of stars and clusters are also characterized by scale-free power-law distributions for luminosity and radius  \citep{ivanov05, 2005A&A...443...79B}. 

The radius distribution of clusters in M51 is well described by a power-law \citep{bastian05}. For the interstellair gas clouds there is a clear relation between radius and mass of the form $M\propto R^{\sim 2}$, which is found back for stellar complexes \citep{2001AJ....121..182E, 2005A&A...443...79B}. This relation breaks down at the level of individual clusters, for which observations are more consistent with no relation between mass and radius at all (e.g. \citealt{1999AJ....118..752Z, 2004A&A...416..537L}). For the embedded clusters in the Milky Way this relation is still present \citep{2007prpl.conf..361A}, suggesting that the relation gets wiped out by a combination of effects in the early stages of evolution (e.g. gas removal, dynamical mixing, tidal truncation etc.). However, the range of masses of these clusters span only a few orders of magnitude, which makes it dangerous to extrapolate this result to all masses. Several scenarios have been introduced to explain the scatter relation between cluster radius and mass, involving a mass dependent star formation efficiency  \citep{2001AJ....122.1888A}  or the early evolution were gas expulsion plays a role \citep{2006MNRAS.373..752G}. Still, a concluding explanation can not be given yet. 

Star clusters with masses in excess of a few times $10^6\,\msun$ tend to be slightly larger than the typical radii found for young clusters and globular clusters ($\sim3\,$pc). The mass-radius relation for these super massive clusters  is well described by the Faber-Jackson relation for elliptical galaxies, when extrapolated down to these masses  (\citealt{2005ApJ...627..203H, 2006A&A...448.1031K} and in these proceedings). When clusters form through merging of multiple clusters into a single massive object, the final radius is somewhat larger than the typical radii ($\sim 3\,$pc) of clusters of lower mass  \citep{2005ApJ...630..879F}. This points at a difference between the formation process  of clusters with $M\lesssim10^6-10^7\,\msun$ and those with $M\gtrsim10^6-10^7\,\msun$ \citep{2006A&A...448.1031K}.

\section{Early evolution }
\label{sum:sec3}

The early evolution (first $\sim10\,$Myr) of clusters is dictated predominantly by the removal of left-over gas (not used for star formation) by stellar winds from early-type stars. This removes a significant fraction of the binding energy of the embedded cluster on a time-scale shorter than, or comparable to, the  crossing time of stars in the cluster. This causes the cluster to expand, lose mass or even completely dissolve (e.g. \citealt{1980ApJ...235..986H, 1997MNRAS.286..669G, 2001MNRAS.321..699K, 2001MNRAS.323..988G}). The simulations of clusters including residual gas expulsion  generally assume a spherically symmetric gas configuration and instantaneous gas removal, resulting in an unbound cluster whenever the star formation efficiency is less than 50\%, i.e. when the total gas mass that is expelled equals the mass of the stars.  Detailed hydrodynamic calculations, taking into account the clumpiness of the gas, show that only a small fraction of the gas is accelerated to high velocities, escaping through irregular outflow channels, causing a larger fraction of stars to remain bound (\citealt{2005MNRAS.358..291D} and Clarke in these proceedings). From observations it appears that some clusters remove their primordial gas more efficiently than others. For example, the extremely  young ($1\pm1\,$Myr, \citealt{2004AJ....128..765S}) Galactic cluster NGC~3603 has cleared its inner region, while NGC~346 in the Small Magellanic Cloud is still surrounded by gas, even in its core,  whereas its age is $3\pm1\,$Myr (\citealt{2007AJ....133...44S} and Nota in these proceedings).

Rapidly expanding clusters have a wide range of implications on the evolution of galactic discs \citep{2005tdug.conf..629K}, the CIMF \citep{2003MNRAS.338..673B} and the early disruption, or {\it infant mortality}, of clusters \citep{2003ARA&A..41...57L}. In addition, estimates of dynamical masses at young ages ($\lesssim30\,$Myr) are affected, since these rely on the assumption that the cluster is in virial equilibrium, which is not the case right after gas expulsion \citep{2006MNRAS.373..752G}.

Several observational confirmations of the infant mortality scenario have become available recently. The most direct one comes from the age distribution (\dndt) of (young) clusters. For  the solar neighbourhood \citet{2003ARA&A..41...57L} find that there is a steep drop in \dndt\ going from the embedded ($\lesssim\,3$Myr) clusters to the young clusters that have expelled their gas ($\gtrsim\,3$Myr), corresponding to $\sim90\%$ infant mortality. This value was confirmed independently, by determining the mean formation rate of bound clusters in the last few Gyrs, correcting for various disruption processes,  and comparing this to the total star formation rate following from field stars (\citealt{2006A&A...455L..17L} and in these proceedings).

From the age distribution of (optically detected) clusters, \citet{2005ApJ...631L.133F} and \citet{2005A&A...443...79B} derive similar values for the infant mortality rates for the clusters in the Antennae galaxies and M51, respectively.  A novel approach is presented by \citet{pellerin} (also in these proceedings). They study groupings of individual stars of different spectral type in nearby (few Mpc) spiral galaxies using HST/ACS data.  The O stars are strongly clustered, while the B stars are more equally spread out over the disc, suggesting that infant mortality works on time-scales comparable to the life-time of O stars. From the surface brightness profiles of slightly resolved clusters, \citet{2006MNRAS.369L...9B} find evidence of escaping stars in the outer halo of young (few Myrs) clusters from high resolution HST/HRC imaging.

\section{The evolution of the (globular) cluster mass function}
\label{sum:sec4}
Larsen (in these proceedings) highlights some of the standing problems in our understanding of the globular cluster mass function (GCMF). The general picture is that clusters form from a universal power-law CIMF with index $-\alpha=-2$, as is observed for young clusters in several galaxies (e.g.  \citealt{1999ApJ...527L..81Z, 2003MNRAS.343.1285D, 2003A&A...397..473B}). Through dynamical evolution, the low-mass clusters are  preferentially destroyed, causing the GCMF to {\it turn-over} (e.g. \citealt{1990ApJ...351..121C, fz01}).  Although the general idea is quite established, there are still several issues to be resolved. First, if dynamical evolution removes low-mass clusters, then the location of the turn-over in the GCMF, i.e. the turn-over mass, will depend on the strength of the tidal field. This implies a decrease of the turn-over mass with galactocentric distance  (\rg), because the strength of the tidal field decreases with \rg.

However, from observations  it follows that the turn-over mass is almost universal among different galaxy types and independent of \rg\ (e.g. \citealt{richtler03}). Solutions to overcome this problem have been introduced in analytical models, such as including a strongly increasing velocity anisotropy with \rg\ \citep{fz01}. Observations of globular cluster kinematics have shown that these assumptions are not realistic (e.g. \citealt{2001ApJ...559..828C} for the M87 clusters).

The term ``turn-over" can be misleading. The $\dr N/\dr \log M$ distribution, or the magnitude distribution (since the magnitude scales with the logarithm of the mass, assuming a constant $M/L$-ratio), is peaked. A $\dr N/\dr M$ representation of the GCMF is flat on the low-mass end and on the high-mass end it approaches the (initial?) power-law distribution with index $-2$ (see Fig.~2 of Larsen in these proceedings). 

If the disruption time of clusters (\tdis) depends on their mass as a power-law with index $\gamma$, then the low-mass end of a {\it single-age population} evolves to a power-law with index $1-\gamma$ \citep{2005A&A...441..117L}. A flat $\dr N/\dr M$ distribution is thus achieved when $\gamma=1$. It is worth noting that the final index is independent of the {\it initial} index $\alpha$, implying that the low-mass end of the globular cluster MF reflects only disruption and not formation. This holds of course only if the initial index $-\alpha$ is smaller than $1-\gamma$. The evolution of GCMFs that are bell-shaped initially are discussed by \citet{2000MNRAS.318..841V}.

For a {\it multi-age population} which has formed with a continuous rate and for which the age spread is much larger than the typical \tdis, the index of the $\dr N/\dr M$ distribution approaches a value of $\gamma-\alpha$ \citep{2003MNRAS.338..717B} (BL03), which is $-1$ in the case of $-\alpha=-2$ and $\gamma=1$, i.e.  steeper by an index 1 as compared to the  single-age case. This type of multi-age populations were studied by BL03 and they find a mean index of the mass function of $\simeq-1.4$, implying $\gamma\simeq0.6$, for the cluster populations in the SMC, M33, M51 and the solar neighbourhood.

\section{Some notes on $\gamma$}
\label{sum:sec5}

The value of $\gamma=1$ is needed to get the flat  \dndm\ distribution of the GCMF on the low-mass end, while BL03 find $\gamma\simeq0.6$ from the \dndm\ distribution of clusters with ages up to a few Gyrs. Did globular clusters evolve differently than their younger counterparts? Or did globular clusters not form with the same initial mass function as clusters now? 

There is theoretical support for a value of $\gamma=0.6$. The arguments are as follows: \citet{baumgardt2001} showed that  \tdis\ depends on the relaxation time (\trel) and the crossing time (\tcr) of stars in the cluster as $\tdis\simeq t^x_{\rm rh}\, t^{1-x}_{\rm cr}Ê$ which, with the expressions for \trel\ and \tcr\ from \citet{1987degc.book.....S}, can be reduced to 
\begin{equation}
\tdis\propto\left[\frac{N}{\ln \Lambda}\right]^x\left[\frac{\rh^3}{\Mc}\right]^{1/2},
\label{eq:tdis}
\end{equation}
 with $\Lambda$ the Coulomb logarithm which scales with $N$ as $\Lambda\simeq0.1N$ \citep{1987degc.book.....S} and \rh\ the half-mass radius of the cluster.  \citet{2003MNRAS.340..227B} verified this relation using detailed $N$-body simulations of clusters dissolving in the Galactic tidal field and they found that $x\simeq0.75$. They assumed that clusters are initially in tidal equilibrium with the Galaxy, implying a constant cluster density, i.e. $\rh\propto\Mc^{\lambda}$ with $\lambda=1/3$, at a given galactocentric distance. This reduces Eq.~\ref{eq:tdis} to $\tdis\propto\left[{N}/{\ln \Lambda}\right]^{0.75}$. The same scaling of \tdis\ with $N$ was found by \citet{1997MNRAS.289..898V}. \citet{2005A&A...429..173L} showed that this relation can be well approximated by $\tdis\propto N^{0.6}$, agreeing with the empirical findings of BL03. The scaling of $\Lambda$ with $N$ is often ignored by taking $\ln\Lambda$ constant (e.g. \citealt{fz01}). Combined with the assumption of a constant cluster density, i.e. $\lambda=1/3$,  then this results in a linear scaling of \tdis\ with $N$ (Eq.~\ref{eq:tdis}), equivalent to $\gamma=1$, which is required to model the low-mass end of the GCMF \citep{2006ApJ...650..885W}. However, this value of $\gamma$ does not follow from  recent $N$-body simulations, but from the  assumptions that \tdis\ scales linearly with \trel, a contant $\Lambda$ and a constant cluster density.
 
The assumption of \citet{2003MNRAS.340..227B} and \citet{1997MNRAS.289..898V} that clusters start tidally limited, i.e. $\lambda=1/3$, is also not realistic. From recent observations it follows that the relation between $\rh$ and $\Mc$ is much shallower, i.e. $\lambda\simeq0.1$ (e.g. \citealt{1999AJ....118..752Z, 2004A&A...416..537L, bastian05}). With such a mass-radius relation, the massive clusters are not filling their  ``Roche lobe". This situation was considered in $N$-simulations by \citet{tf05}. They modeled the disruption of clusters in weak tidal fields, where the tidal radius of the clusters is smaller than the Roche lobe radius. They find that in those cases $x\simeq0.9-1.0$, i.e. almost as if the cluster is evolving in isolation (then $x\equiv1$, \citealt{1987degc.book.....S}). Hence, $x$ depends on $\lambda$. Lets consider the case of a realistic mass-radius relation, i.e. $\lambda=0.1$ \citep{2004A&A...416..537L}.  Assume a constant mean stellar mass, i.e. $\Mc\propto N$ and lets approximate ${N}/{\ln \Lambda}$ by $N^{0.85}$, which is a good approximation for $N>>N/\Lambda\simeq10$. This, combined with $x\simeq1$ and Eq.~\ref{eq:tdis} results in $\gamma\simeq0.5$. When clusters are disrupted by external perturbations (e.g. disc/bulge shocks, encounters with giant molecular clouds or spiral arms), \tdis\ scales with the cluster density (\rhoc) (e.g. \citealt{1987degc.book.....S, 2006MNRAS.tmp..808G, 2007astro.ph..1136G}), which combined with $\lambda=0.1$ results in $\tdis\propto\Mc^{0.7}$, again a value for $\gamma$ that is smaller than $1$. 

In neither of the cases described above we get $\gamma=1$ needed for the low-mass end of the GCMF. Hence not only the universality of the turn-over mass, but also the shape of the GCMF can not be explained by the standard picture where clusters start with a power-law CIMF with index $-2$ and evolve dynamically. The three predicted values for $\gamma$ are quite close to the observationally determined  value of $\gamma=0.62\pm0.06$ (BL03).

\begin{acknowledgements}
It is a great pleasure to thank Henny for 4 very pleasant years  as a PhD student in his group! His analytical way of thinking and working have helped me many times when trying to get insight in a scientific problem. This is an experience that I share with many attendants at this conference. I would like to thank the organizers of this conference for a very enjoyable meeting which successfully brought together groups of different disciplines. 
\end{acknowledgements}

\end{document}